\documentclass[aps,pra,twocolumn,groupedaddress,amsmath]{revtex4-1}

\usepackage{graphicx}
\begin{document}


\title{Calculation of ground state energy of harmonically confined two dipolar fermions}


\author{Amit K. Das$^{a}$ and Arup Banerjee$^{b}$}
\email[]{banerjee@rrcat.gov.in}
\affiliation{$^a$Laser Material Processing Division, Raja Ramanna Centre for Advanced Technology, Indore 452013, India \\
$^b$BARC Training School at RRCAT, Raja Ramanna Centre for Advanced
Technology, Indore 452013, India and \\
Homi Bhabha National Institute, Indore 452013, India}


\date{\today}

\begin{abstract}
We calculate the ground state energies of a system of two dipolar fermions trapped in a harmonic oscillator potential. The dipoles are assumed to be aligned parallel to each other. We perform the calculations of ground state energy as a function of strength of interaction between two fermions by employing variational method with Hylleraas-like explicitly correlated wave function. Furthermore, we perform calculations of ground state energy within Hartree-Fock approximation and the magnitude of correlation energy is estimated by subtracting these results from the corresponding wave function based results. We also carry out calculations of ground state energies within the realm of density functional theory by using recently reported expressions for exchange and correlation energies under local density approximation. By comparing correlated wave function based results with those obtained using density functional theory approach we examine the role of fermion-fermion correlation and assess the accuracy of local density approximation based expression for the correlation energy functional.
\end{abstract}


\maketitle


\section{Introduction}
Recently, several studies on the static and dynamic properties of dipolar Fermi gas have been reported in the literature \cite{rzazewaski2001,rzazewaski2003,pu,bohn,pelster1,pelster2,yin,adhikari}. For general review on both dipolar bosons and fermions we refer the readers to Ref. \cite{baranov}. The characteristic properties of dipolar Fermi gas are distinctly different from that of an electron gas or usual Fermi gas made of non-dipolar atoms due to the anisotropic and long-range nature of the interaction between the dipoles. The dipolar Fermi gases have been realized experimentally by using various atomic species with large magnetic moment. For example, isotopes of $^{53}$Cr, $^{173}$Yb, and $^{161}$Dy  atoms with large magnetic moment of 6$\mu_{B}$ ($\mu_{B}$ denotes Bohr magnetron), 3$\mu_{B}$, and around 10$\mu_{B}$, respectively have been cooled down to reach quantum degeneracy \cite{gorceix,takahashi,lev}. On the other hand, heteronuclear polar diatomic molecules with large electric dipole moment are promising candidates for realization of dipolar Fermi gases. Recently, quantum degeneracy in a sample of around $4\times 10^{4}$ fermionic $^{40}$K$^{87}$Rb molecules in their rovibrational ground state with permanent electric dipole moment of 0.566 Debye has been achieved \cite{ni1}. Furthermore, the anisotropic characteristic of the dipole-dipole interaction has also been experimentally investigated in these KRb samples \cite{ni2}. 

The successful creation of degenerate dipolar Fermi gas of KRb molecules has given rise to a new class of many-body system and this has rekindled the interests of theorists to look into these systems. The quantum mechanical description of degenerate Fermi gas is more complicated than its bosonic counterpart due to the antisymmetric nature of the many-fermion wave function. Therefore, for theoretical description of many-fermion systems it becomes necessary to invoke some approximations. We note here that most of theoretical studies on dipolar Fermi gases have been carried out either within the semi-classical Thomas-Fermi (TF) approximations \cite{rzazewaski2001}, which neglects effects of both exchange and correlation between the fermions or within the Hartee-Fock (HF) theory \cite{pu,bohn,pelster1,pelster2}, which takes into account the quantum mechanical effect of exchange of identical particles. It is only recently Liu and Yin \cite{yin} reported calculation of correlation energy of homogeneous dipolar Fermi gas by going beyond HF approximation.
They estimated the correlation energy, which is defined as the difference between the true ground state energy and the corresponding energy in the HF approximation using perturbative Brueckner-Goldstone formalism along with Monte Carlo integration to fix the coefficient. This study showed that the ground state energy of a homogeneous dipolar Fermi gas is reduced when correlation effect is included resulting in significant lowering of critical density required for mechanical collapse as compared to the HF case. These authors also employed the expression for the correlation energy to calculate the properties of inhomogeneous trapped dipolar Fermi gas by using the expressions for exchange and correlation energy functionals within local density approximation (LDA).

We wish to note here that for a system consisting of few (two to three) interacting fermions it is possible to construct a fully correlated wave function satisfying antisymmetric property for the calculation of its ground state energy.  For example, in atomic and molecular physics Hylleraas-like correlated wave functions \cite{kolos,szalewicz} with explicit dependence on the inter-electronic coordinates have yielded very accurate results for two-electron helium-like ions and H$_{2}$ molecule \cite{scully}. It should be mentioned here that beside exchange effect a correlated wave function also takes into account the effect of electron-electron correlation arising from the electron-electron interaction through the dependence of wave function on the inter-electronic coordinate. However, both HF approach ( which takes into account the effect of exchange) and semi-classical TF approximation neglect effect of two-particle correlations. Our main objective of this paper is to first calculate ground state energy of a system of two dipolar fermions confined in a harmonic trap by using Hylleraas-like correlated wave function. For the purpose of comparison we also carry out the calculation of ground state energy of a confined two-dipolar fermion system by using HF approach. 

Albeit, the correlated wave function approach takes into account the effect of both exchange and correlation however, the form of the wave function becomes very cumbersome with increasing number of particles making this approach impractical for systems beyond three particles. A formalism which takes into account the effects of both exchange and correlation in many-fermion systems efficiently is  the Density functional theory (DFT) \cite{parr,perdew,engel}. This formalism circumvents the use of complicated many-body wave function of 3N spatial variables ( where N is number of particles), rather works with particle density - a function of just three spatial variables. As a result of this DFT based method turns out to be quite suitable for studying fermionic systems with large number of particles.  Although the theory is exact in principle but the exact forms for exchange and correlation functional in term of density are not known. Therefore, to carry out any DFT based calculations it becomes necessary to use approximate expressions for the exchange and correlation energies.  For electronic systems a tremendous amount of effort has gone into the development of exchange and correlation functionals with increasing degree of accuracy \cite{perdew,engel}. The availability of accurate exchange-correlation functional along with significantly lower computational cost of DFT-based calculations has made DFT an indispensable tool for handling many-electron systems like atoms, molecules, and solids. For electronic systems the simplest and one of the most extensively used forms for exchange and correlation energy functionals are constructed within LDA by replacing the constant electron density in the expressions for exchange and correlation energies of a homogeneous electronic gas by the local density of the inhomogeneous system. 
As mentioned above it is only recently expressions for the exchange and correlation energies of a homogeneous dipolar Fermi gas have been derived \cite{bohn,yin}. The availability of these expressions has motivated us to carry out DFT based calculations of the ground state energy of two harmonically trapped dipolar Fermi particle within LDA and compare these results with the numbers obtained via correlated wave function approach to assess their accuracies.

The remaining paper is organized in the following manner. In section II we present the model and describe various theoretical methods employed to calculate the ground state energy. The results are presented and discussed in section III. The paper is concluded in section IV. 

\section{Model and various methods to obtain ground state energy of two dipolar fermions confined in a harmonic trap}

In this paper we consider a system of harmonically trapped two identical fermionic particles of mass $m$ possessing electric dipole moment $d$, which are polarized along the z-axis. The non-relativistic Schr$\ddot{o}$dinger equation for such a pair of identical dipolar fermions confined in a harmonic oscillator potential with frequency $\omega$ is given by,
\begin{widetext}
\begin{equation}
\left [ -\frac{1}{2}{\vec{\nabla}}_{1}^{2} - \frac{1}{2}{\vec{\nabla}}_{2}^{2} + \frac{1}{2}\left (r_{1}^{2} + r_{2}^{2}\right ) + v_{dd}({\bf r}_{1},{\bf r}_{2})\right ]\Psi({\bf r}_{1},{\bf r}_{2}) = E\Psi({\bf r}_{1},{\bf r}_{2}).
\label{schrodinger1}
\end{equation}
\end{widetext}
The above equation has been written in the units of length $l_{0}$ ($l_0 = \sqrt{\hbar/m\omega}$) and energy $\epsilon_{0}$ ($ \epsilon_{0}=\hbar\omega$) of cofining harmonic oscillator potential. The potential $v_{dd}$ in Eq. (\ref{schrodinger1}) represents the dipole-dipole interaction between the fermions,
\begin{equation}
v_{dd}({\bf r}_{1},{\bf r}_{2}) = \frac{3a_{dd}}{l_0}\frac{(1-3\cos^2\theta)}{r_{12}^3},
\label{dipoleScaled}
\end{equation}
where $3a_{dd}=md^2/4\pi\epsilon_0\hbar^2$ is the effective interaction strength and $\theta$ is the angle between the vector $\bf{r_{12}} = \bf{r_1} - \bf{r_2}$ and the direction of the dipoles, which is chosen to be along the z-axis. 

To calculate ground state energy of trapped dipolar two-fermion system we need to solve the eigenvalue equation given by Eq. (\ref{schrodinger1}). For this purpose we apply Rayleigh-Ritz variational method with an appropriate trial wave function for $\Psi ({\bf r_1, r_2})$ . To describe the correlation effect accurately we choose to use a trial wave function for the two-fermion system with explicit dependence on the fermion-fermion separation $r_{12}$. 
To this end we employ  wave function of the form (S-symmetry)
\begin{widetext}
\begin{equation}
\Psi ({\bf r_1, r_2})= N\left[ \exp(-\alpha r_1^2-\beta r_2^2)+\exp(-\alpha r_2^2-\beta r_1^2)\right]\left[1-\exp(-\lambda r_{12})\right],
\label{hylleraas}
\end{equation}
\end{widetext}
where $N$ is normalization constant. The total wave function is product of $\Psi ({\bf r_1, r_2})$ and the the spinor wave function $\chi (1,2)$. As spatial part ($\Psi ({\bf r_1, r_2})$) is symmetric with respect to the interchange of coordinates the spinor part must be antisymmetric to ensure the antisymmetric nature of the total wave function. In the above equation (Eq. (\ref{hylleraas}))  the three nonlinear coefficients $\alpha$, $\beta$, and $\lambda$ are variational parameters, which are determined by minimizing the ground state energy as described below.  We note here that keeping in mind the harmonic nature of the confining potential we choose gaussian functions  involving the coordinate variables $r_1$ and $r_2$. The form of the correlation function is motivated by the fact that unless the wave function goes to zero as $r_{12} \to 0$, the integral for the interaction energy diverges. Also because of the finite size of the dipoles it is physically not possible for the dipoles to come within distance $1/\lambda$ . 

The Rayleigh-Ritz variational method for solving Eq. (\ref{schrodinger1}) involves finding the minimum value of the following energy functional of trial wave function $\Psi$
\begin{widetext}
\begin{equation}
E[\Psi ] = \frac{\int\int\Psi^{*}({\bf r}_{1},{\bf r}_{2})\left [-\frac{1}{2}{\vec{\nabla}}_{1}^{2} - \frac{1}{2}{\vec{\nabla}}_{2}^{2} + \frac{1}{2}\left (r_{1}^{2} + r_{2}^{2}\right ) + v_{dd}\right ]\Psi({\bf r}_{1},{\bf r}_{2})d{\bf r}_{1}d{\bf r}_{2}}{\int\int\Psi^{*}({\bf r}_{1},{\bf r}_{2})\Psi({\bf r}_{1},{\bf r}_{2})d{\bf r}_{1}d{\bf r}_{2}}.
\label{energyfunc1}
\end{equation}
\end{widetext}
By substituting the wave function given by Eq. (\ref{hylleraas}) in the above equation and using the coordinate system involving $r_1, r_2$ and $r_{12}$, it can be reduced to sum of integrals of the form
\begin{equation}
\int_{0}^{\infty} dr_{1}\int_{0}^{\infty} dr_{2} \int_{|r_{1} - r_{2}|}^{r_{1} + r_{2}} dr_{12} f(r_1, r_2, r_{12})
\end{equation}
by using the differential volume element as \cite{murphy}
\begin{equation}
dV = 8\pi^{2}r_{1}r_{2}r_{12}dr_{1}dr_{2}dr_{12},
\end{equation}
with $f(r_1, r_2, r_{12})$ being a function of $r_1, r_2$ and $r_{12}$. The energy functional is minimized by varying three nonlinear parameters $\alpha$, $\beta$, $\lambda$ to obtain the ground state energy. 

Having described the correlated wave function based approach for the calculation of ground state energy we proceed with a brief description of the method employed to calculate the ground state energies of dipolar Fermi system within the realm of DFT. In accordance with the Hohenberg-Kohn theorem \cite{kohn} of DFT the ground state energy of a dipolar two-fermion system can be uniquely expressed as a functional of the fermion density $\rho$:
\begin{equation}
E[\rho] = T[\rho] + V_{conf}[\rho] + E_{dd}[\rho] + E_{xc}[\rho],
\label{dft1}
\end{equation}
where $T[\rho]$ and $V_{conf}[\rho]$ are the non-interacting kinetic and confinement energies respectively. For a two-fermion system considered in this paper $T[\rho]$ and $V_{conf}[\rho]$ are given by,

\begin{equation}
T[\rho] = -\frac{1}{2}\int \sqrt{\rho(\bf{r})}\nabla^2\sqrt{\rho(\bf{r})}d{\bf r},
\label{kinetic_energy}
\end{equation}

and

\begin{equation}
V_{conf}[\rho] = \frac{1}{2}\int r^2\rho({\bf r}) d{\bf r}.
\label{conf_energy}
\end{equation}
The third term $E_{dd}[\rho]$ is the dipolar counterpart of the Hartree energy representing classical electrostatic energy between two dipoles and it is given by 
\begin{equation}
E_{dd}[\rho] = \frac{1}{2}\int v_{dd}({\bf r}_1,{\bf r}_2)\rho({\bf r}_1)\rho({\bf r}_2)d{\bf r}_1 d{\bf r}_2,
\label{dd}
\end{equation}
where $v_{dd}({\bf r}_1,{\bf r}_2)$ is the dipole interaction potential as defined in Eqs. (\ref{dipoleScaled}). The last term $E_{xc}[\rho]$ represents the the exchange-correlation (XC) energy arising purely due to quantum mechanical nature of two trapped fermions. Furthermore,  the second part of Hohenberg-Kohn theorem provides the energy variational principle according to which the ground state density can be obtained by making the energy functional (Eq. (\ref{dft1})) stationary. As mentioned above, in general the exact forms for the exchange and correlation energy functionals are not known and thus for implementing DFT based calculations one needs to use approximate forms for these energy functionals. In the present paper we use the expressions for exchange and correlation energies of homogeneous dipolar Fermi gas derived recently in Ref. \cite{pu,yin} within LDA by replacing the uniform fermion density by the local density $\rho ({\bf r})$ of the inhomogeneous system.  In accordance with Ref. \cite{yin} we employ following expressions for the exchange ($E_{x}$) and correlation ($E_{c}$) energy functionals within LDA:
\begin{equation}
E_{x}[\rho] = - \frac{4}{45}(6\pi^2)^{1/3}(3a_{dd})^2\int \rho^{7/3}({\bf r})d{\bf r},
\label{exchange}
\end{equation}
\begin{equation}
E_{c}[\rho] = - \frac{2}{3}(3a_{dd})^2\int \rho^{7/3}({\bf r})d{\bf r}.
\label{correlation}
\end{equation}
In Ref \cite{yin} the above forms for the exchange and correlation energy functionals have been applied to study the mechanical collapse of trapped fermionic KRb polar molecules. 

Once again we make use of variational principle as guaranteed by Hohenberg-Kohn theorem to carry out calculation of the ground state energy and the corresponding density. 
To this end we need to make a judicious choice for the form of trial ground state density $\rho ({\bf r})$. Keeping in mind the harmonic trapping potential and the anisotropic nature of the dipole-dipole interaction term we choose following form for the trial density for the ground state:
\begin{equation}
\rho(r) = 2|\phi(r)|^2,
\label{density}
\end{equation}
with
\begin{equation}
\phi(r)=\sum_p C_p \  \exp \left[-(w_p x^2+w_p y^2+\gamma_p z^2)\right].
\label{basis}
\end{equation}
In Eqs. (\ref{basis}) $w_p$ and $\gamma_p$ are the variational parameters and the above form of density is also normalized to number of fermions.  The constants $C_p$'s  are such that $\phi(r)$ remains always normalized. With the above choice for the form of the trial density, the kinetic, confinement, and interaction energy terms in the energy functional given by Eq. (\ref{dft1}) can be simplified to,
\begin{widetext}
\begin{equation}
T[\rho] = 2\pi^{3/2}\sum_{pq} \frac{C_pC_q}{(w_p+w_q)\sqrt{\gamma_p+\gamma_q}} \left[ 2w_q+\gamma_q-\frac{2w_q^2}{(w_p+w_q)}-\frac{\gamma_q^2}{(\gamma_p+\gamma_q)} \right],
\label{kinetic}
\end{equation}

\begin{equation}
V[\rho] = 2\pi^{3/2}\sum_{pq} \frac{C_pC_q}{4(w_p+w_q)\sqrt{\gamma_p+\gamma_q}} \left[ \frac{2}{(w_p+w_q)}+\frac{1}{(\gamma_p+\gamma_q)} \right],
\label{potential}
\end{equation}

\begin{equation}
E_{dd}[\rho] = -\sum_{pqrs} \frac{8\pi^{5/2} C_p C_q C_r C_s}{3(w_p+w_q+w_r+w_s)\sqrt{(\gamma_p+\gamma_q + \gamma_r+\gamma_s)}}\frac{3a_{dd}}{l_0}f(\kappa),
\label{Edd}
\end{equation}
\end{widetext}

where 
\begin{equation}
\kappa^2 = \frac{(\gamma_p+\gamma_q)(\gamma_r+\gamma_s)(w_p+w_q+w_r+w_s)}{(w_p+w_q)(w_r+w_s)(\gamma_p+\gamma_q+\gamma_r+\gamma_s)},
\end{equation}
and

\begin{equation}
f(\kappa) = \begin{cases} \frac{1+2\kappa^2}{1-\kappa^2}-\frac{3\kappa^2 \tanh^{-1}\sqrt{1-\kappa^2}}{(1-\kappa^2)^{3/2}}, \mathrm{ if } \  |\kappa|<1 \\ 0, \ \mathrm{if } |\kappa|=1 \\ -\frac{1+2\kappa^2}{\kappa^2-1}+\frac{3\kappa^2 \tan^{-1}\sqrt{\kappa^2-1}}{(\kappa^2-1)^{3/2}}, \  \mathrm{ if } |\kappa|>1\end{cases}.
\label{f_k}
\end{equation}

Using above expressions (Eqs. (\ref{exchange}) - (\ref{f_k})) we minimize the total energy by varying the parameter $w_p$'s, $\gamma_p$'s and $C_p$'s to obtain ground state density and the corresponding energy. The the number of basis functions ($p$) in Eq. (\ref{basis}) is fixed by checking the convergence of the ground state energy with increasing number of basis functions.

In order to assess the accuracy of LDA correlation energy functional employed by us (Eq. \ref{correlation}) we first estimate the correlation energy component from the total energy obtained via wave function based approach. For this purpose we use HF based definition of the correlation energy. As mentioned before the correlation energy is defined as the difference between the exact ground state energy and the corresponding energy in the HF approximation.  Therefore, in order to estimate correlation energy we also carry  out calculations of the ground state energy $E_{HF}$ within HF approximation. The correlation energy is then obtained by subtracting $E_{HF}$ from the corresponding result obtained with correlated wave function approach.  The calculations within HF approximation can be performed employing the procedure described above for DFT based calculations by substituting  $E_{c} = 0$ and an exact expression for the exchange energy for two fermion system given by 
\begin{equation}
E_{x} = -\frac{1}{2}E_{dd}.
\end{equation}
In the next section we present the results of our calculations obtained by applying various approaches mentioned above and compare them to test their accuracies. Before proceeding to the next section it should be mentioned here that in general the correlation energy obtained within DFT differs from the above definition due to the presence of kinetic energy component in the DFT based definition of the correlation energy. However, for two-particle system considered in this paper the contribution of kinetic energy is zero and two definitions of correlation energy coincide \cite{engel}.

\section{Results and Discussions}  
The correlated wave function based variational calculation for determining the ground state energy has been performed using GNU Scientific Library (GSL) routines  \cite{gsl}. The variational integrations were carried out using VEGAS algorithm of Monte-Carlo multidimensional integration method and the energy minimization calculations have been  performed using Nelder-Mead algorithm as implemented in GSL. In Table \ref{tabone} we present the results for the ground state energies obtained by employing correlated wave function for different values of interaction strength determined by the dimensionless $\eta = 3a_{dd}/l_{0}$ ranging from value 0.0 to 1.0. From this Table we observe that for $\eta = 1.0$ the ground state energy ($E_{GS}$) of a harmonically confined two dipolar fermions is $2.715$. Furthermore, on decreasing the strength of the dipolar interaction, the ground state energy increases gradually and it correctly approaches the energy of two non-interacting harmonically confined fermions ( $E_{non-int} = 3.0$) as the interaction is completely switched off. We note here that the ground state energy of two interacting dipolar fermions confined in a harmonic oscillator potential is less than its non-interacting counterpart which clearly  indicates that the dipole-dipole interaction is on the average attractive in nature. 

\begin{table}
\caption{\label{tabone}Ground state energy as a function of strength ($\eta$) of the dipolar interaction as calculated from the correlated variational function.} 
\begin{ruledtabular}
\begin{tabular}{c c}                          
$\eta=3a_{dd}/l_0$  &  Ground state energy \\
\hline 
$1.0$ & $2.715$ \\
$0.9$ & $2.750$ \\
$0.8$ & $2.788$ \\
$0.7$& $2.806$ \\
$0.6$& $2.838$ \\
$0.5$& $2.866$ \\
$0.4$& $2.896$ \\
$0.3$& $2.922$ \\
$0.2$& $2.950$ \\
$0.1$& $2.975$ \\
$0.0$& $3.000$ \\
\end{tabular}
\end{ruledtabular}
\end{table}

Having obtained the results for the ground state energies by using the correlated wave function given by Eq. (\ref{hylleraas}) we now present the corresponding results obtained via DFT and HF based approaches. These calculations have been performed by employing finite basis sets consisting of gaussian functions (Eq. (\ref{basis})). Therefore, first we check the convergence of our results by increasing number of basis functions. The results for the ground state energy obtained by employing DFT and HF based methods for interaction strength $\eta = 1.0$ with increasing number of basis functions are presented in Table \ref{tabtwo}. We find that both DFT and HF based results converge with the inclusion of three or more basis functions. Therefore all the DFT and HF based results presented below have been obtained by performing calculations using only three basis functions. To accomplish the main aim of this paper we now proceed with the comparison of the results obtained by wave function based approach with those obtained via DFT and HF methods.

\begin{table*}
\caption{\label{tabtwo}Ground state energy as a function of number of basis function in Eqs. (\ref{basis}).} 
\begin{ruledtabular}
\begin{tabular}{c c c}                          
Number of basis function & Ground state energy from DFT & Ground state energy from HF  \\
\hline
$1$& $--$& $2.992$\\
$2$& $2.763$& $2.990$\\
$3$& $2.755$& $2.990$\\
$4$& $2.754$& $2.990$\\
$5$& $2.755$& $2.990$\\
\end{tabular}
\end{ruledtabular}
\end{table*}

To this end in Figure 1 we display the ground state energies obtained by employing DFT and HF based approaches along with the ones calculated by using Hylleraas-like correlated wave function given by Eq. (\ref{hylleraas})  versus interaction strength $\eta$. First of all it can be seen from Figure 1 that both HF and DFT methods overestimate the ground state energy in comparison to the corresponding wave function based values. Among three approaches the wave function based method yields lowest values for the ground state energy which is consistent with the fact that this method takes into account both effects of exchange and correlation most accurately. Moreover, note that the ground state energy obtained by employing wave function and DFT based approaches display similar trend with respect to the variation of interaction strength. In particular we observe from Figure 1 that ground state energies obtained by employing wave function and DFT base methods decrease from non-interacting ($\eta = 0$) value of $E_{GS} = 3.00$ as $\eta$ is increased. On the other hand, the ground state energy calculated by HF method remains almost constant at a value of about $3.00$ for the whole range of interaction strength considered in this study. This behaviour clearly indicates that for dipolar fermions effect of correlation, which is not taken into account in HF method, is quite important and it contributes significantly to the ground state energy of the system. We note here that although DFT based method overestimates the ground state energy as compared to the ones obtained through correlated wave function approach nonetheless it performs better than HF method in predicting the variation of energy with the interaction strength. This is because, correlation energy is taken into account in DFT, albeit in an approximate way. Also it can be noted from Figure 1 that for stronger interaction ($\eta \sim 1$), the difference between the ground state energies calculated by employing wave function and DFT based methods decreases. From this result we conclude that LDA based expressions for the exchange and correlation energies of dipolar fermions \cite{yin} are capable of yielding sufficiently accurate results for the ground state energy of a harmonically confined two dipolar fermions. Therefore, DFT based method with exchange and correlation within LDA is expected to be suitable for calculations of ground state properties of dipolar Fermi gas with large number of particles generally achieved in experiments for which wave function based methods are prohibitive.

\begin{figure}
\includegraphics{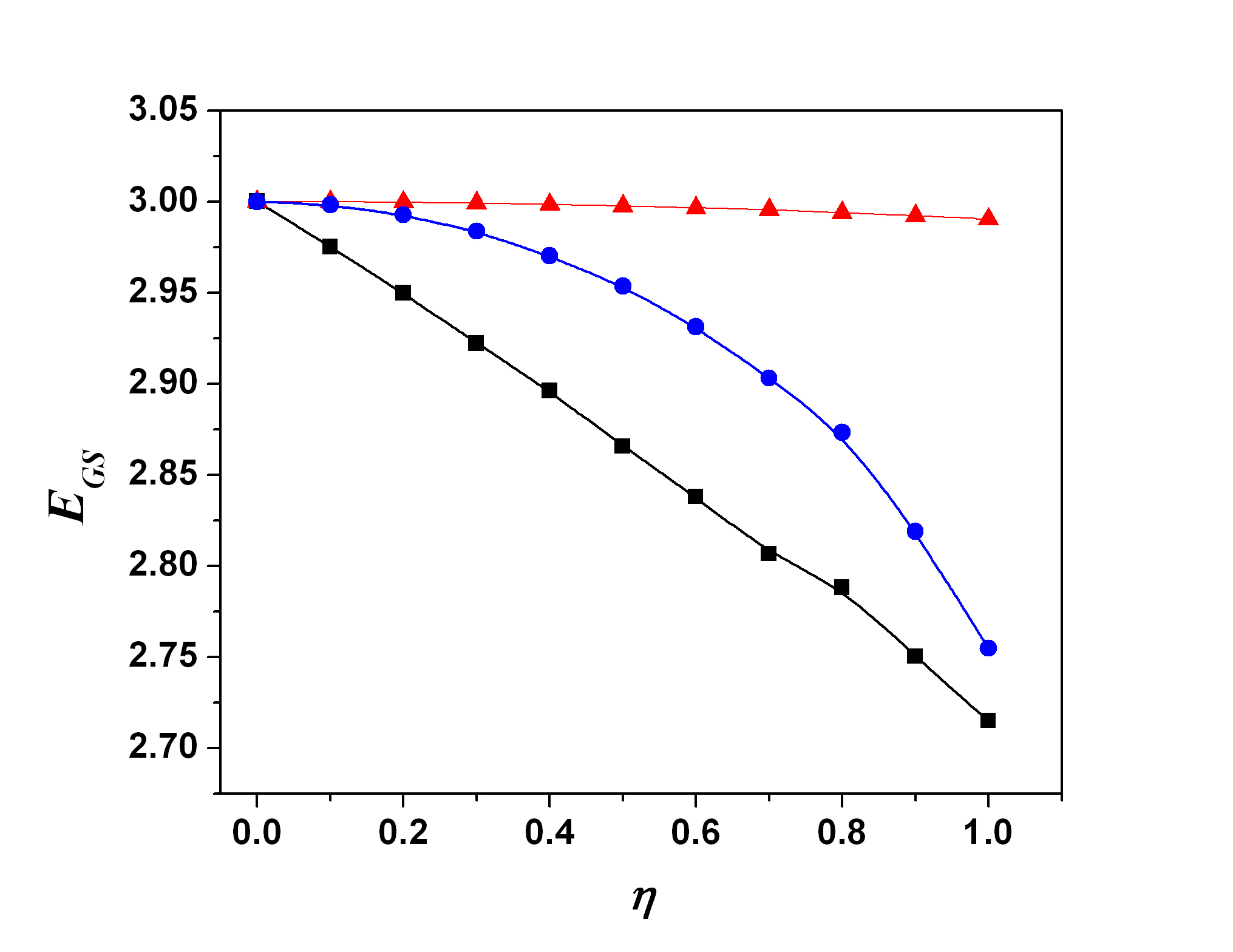}
\caption{The variation of ground state energy calculated by HF (red), DFT (blue) and correlated wavefunction (black) as a function of interaction strength $\eta=3a_{dd}/l_0$. All numbers are in harmonic oscillator unit.}
\end{figure}

Moreover, from Figure 1 and the above discussion it is clear that the contribution of correlation energy to the total energy is quite significant and it increases with increase in the interaction strength. Keeping this in mind we now assess the accuracy of correlation energy functional proposed by Liu and Yin \cite{yin} within LDA. To this end   we compare the correlation energies as calculated by DFT based method with the ones obtained from the explicitly correlated wave function. The contribution of correlation energy ($E_{c}$) obtained by these two methods are displayed in Figure 2 as a function of the strength of dipolar interaction $\eta$. We observe from this Figure that these two methods yield same sign (negative) for the correlation energy and display similar trend with respect to variation of interaction strength $\eta$. However, the LDA based expression for correlation energy underestimates the magnitude of the correlation energies as compared to those calculated from Hyllleraas-like wave function approach for weaker strength of interaction. As the strength of interaction is increased the contribution of correlation energy obtained from DFT based method fast approaches to that calculated from wave function based method and for $\eta \geq 1.00$  the magnitude of correlation energy is overestimated by the LDA based expression.

\begin{figure}
\includegraphics{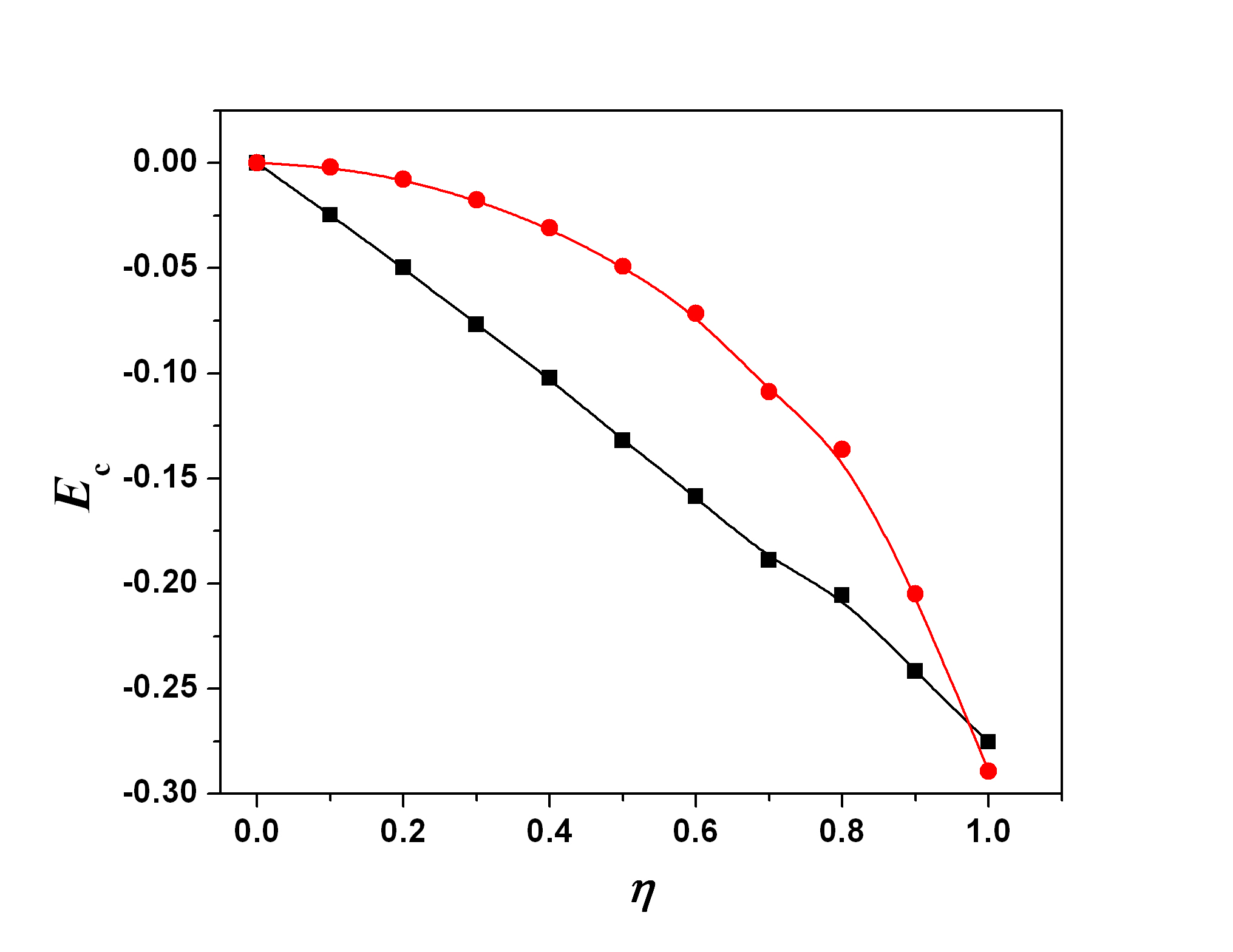}
\caption{The variation of correlation energy calculated by correlated wavefunction (black) and DFT (red) as a function of interaction strength $\eta=3a_{dd}/l_0$. All numbers are in harmonic oscillator unit.}
\end{figure}

Having discussed the general behaviour of total energy and the contribution of correlation energy to it we now feel it is worth here to discuss the energetics for the case of recently achieved  degenerate sample of fermionic KRb molecules with very large value of dipole moment  \cite{ni2}. For this system the value of $3a_{dd}$ is about $6000 a_B$, where $a_B$ is the Bohr radius \cite{Lahaye}. When converted to harmonic oscillator unit with a typical value of oscillator length $l_{0} = 1\times 10^{-6}$m, this corresponds to a strength of the dipole interaction $\eta = 0.317$. For this value of the interaction strength, the ground state energy calculated from correlated wave function comes out to be $E_{GS} = 2.917$. On the other hand, the DFT based result within LDA for ground state energy at this interaction strength is found to be $2.981$, which is around 2$\%$ higher than the corresponding wave function based number. For the same strength of interaction, the HF based calculation yields $E_{GS} = 2.999$. Thus the contribution of correlation energy in the total ground state energy obtained by wave function based approach is found to be -0.082. On the other hand, the value of correlation energy obtained by employing LDA based expression Eq. (\ref{correlation})  yields $E_{c} = -0.018$, which is less than one fourth of the magnitude of correlation energy estimated by wave function based approach. As mentioned before both wave function and DFT based approaches take effect of particle-particle correlation into account whereas HF approach completely neglects it. The difference between the results obtained via wave function and DFT based approaches clearly reveals that the effect of particle-particle correlation is underestimated by the LDA based expression derived in Ref.\cite{yin}.

\section{Conclusion} 
In this paper we have calculated the ground state energy of a harmonically confined two interacting dipolar fermions by variational method using a Hylleraas-like correlated wave function. We study the variation of ground state energy as a function of strength of interaction between the two dipoles. In order to estimate the magnitude of correlation energy  we have also performed calculation of the ground state energy of dipolar fermions within HF approximation. By employing recently derived expressions for the exchange and correlation energy functionals of homogeneous dipolar gas within LDA we have also calculated DFT based ground state energy of two trapped dipolar fermions. The comparison of wave function based results with the ones obtained via DFT clearly shows that two methods exhibit similar trend in the variation of ground state and correlation energies. However, both the energies are underestimated by LDA based calculations as compared to the results obtained via Hylleraas-like correlated wave function. We have also found that the difference between the results obtained by LDA and wave function based methods decreases with increasing interaction strength and they are very close for $\eta \approx 1.00$.
Our study clearly reveals that for dipolar fermions the contribution of correlation energy is quite significant and correct description of correlation effects going beyond LDA is necessary for accurate evaluation of the ground state energy of the system. Despite its limitation we feel that LDA based method within the realm of DFT is suitable for calculating the properties of confined dipolar Fermi gas containing $10^{4} - 10^{5}$ atoms as achieved in the recent experiments. In order to gain more insight the study presented in this paper can be extended to determine the exact exchange and correlation energy functional and potentials by using the density obtained from the correlated wave function. This will be topic of our future publication.



\begin{acknowledgments}
A. D. thanks Dr. L. M. Kukreja and   A. B. wish to thank Dr. P. K. Gupta for constant encouragement and support and Prof. Manoj K. Harbola for comments and suggestions.  
\end{acknowledgments}


\end{document}